# Executable Trigger-Action Comments

Pengyu Nie, Rishabh Rai, Junyi Jessy Li, Sarfraz Khurshid, Raymond J. Mooney, and Milos Gligoric
The University of Texas at Austin
{pynie@,rrai.squared@,jessy@austin.,khurshid@ece.,mooney@cs.,gligoric@}utexas.edu

## ABSTRACT

Natural language elements, e.g., todo comments, are frequently used to communicate among the developers and to describe tasks that need to be performed (actions) when specific conditions hold in the code repository (triggers). As projects evolve, development processes change, and development teams reorganize, these comments, because of their informal nature, frequently become irrelevant or forgotten.

We present the first technique, dubbed TrigIt, to specify trigger-action todo comments as executable statements. Thus, actions are executed automatically when triggers evaluate to true. TrigIt specifications are written in the host language (e.g., Java) and are evaluated as part of the build process. The triggers are specified as query statements over abstract syntax trees and abstract representation of build configuration scripts, and the actions are specified as code transformation steps. We implemented TrigIt for the Java programming language and migrated 20 existing trigger-action comments from 8 popular open-source projects. We evaluate the cost of using TrigIt in terms of the number of tokens in the executable comments and the time overhead introduced in the build process.

## 1 INTRODUCTION

Natural language elements, such as *todo comments*, are frequently used to communicate among developers [28, 39, 41]. Some of these comments specify that a developer should perform an *action* if a *trigger* holds (e.g., from the Google Guava project: "Remove this guard once lazyStackTrace() works in Java 9", from the Apache Wave project: "Remove this when HtmlViewImpl implements getAttributes", etc.). We consider those comments where the trigger is expressed based on the state of the code repositories and actions require modifications of source or binary code. We call these comments *trigger-action comments*.

Although trigger-action comments are ubiquitous, they are, like other types of comments, written in natural language. Thus, as projects evolve, development processes change, and development teams reorganize, these comments frequently become irrelevant or forgotten. As an example, consider the following comment from the Apache Gobblin project [13]: "Remove once we commit any other classes". This comment, followed by an empty class, was included in revision 38ce024 (Dec 10, 2015) in package-info.java file to force the javadoc tool to generate documentation for an empty package. In revision bbdf320 (Feb 4, 2016) classes were added to the package, but the comment (and the empty class in package-info.java file) remained in the code repository up to this date. Having pending actions, i.e., those actions that should have been done because the triggers are satisfied, and outdated comments may have impact on program comprehension, and maintenance [28, 41]. Additionally, having comments written in an informal way presents a challenge for some software engineering tools, such as refactorings [12, 29, 30], as those tools may not know how to manipulate code snippets and identifiers embedded in comments [38].

We present the first technique, dubbed TrigIt, to specify trigger-action comments as *executable statements*; the triggers are evaluated automatically at each build and actions are taken based on users' specifications. TrigIt introduces a domain specific language (DSL) that can be used to specify triggers and actions over the abstract syntax trees (AST) and build configuration scripts. Specifically, triggers are written as *query statements* over ASTs and build configuration scripts. On the other hand, actions are encoded as *transformation steps* on ASTs. To provide a natural access to the AST elements and improve maintenance, the TrigIt DSL is embedded in the host language. However, the semantics of the language enables the executable trigger-action comments to be evaluated as part of the static program analysis phase.

We implemented TrigIt for Java, such that triggers are written using Java streams and actions are transformation steps of custom ASTs. TrigIt analyzes executable code and modifies either executable code, source code, or neither, depending on users' configuration. This flexibility allows users to enforce execution of actions, for example during testing or during debugging of TrigIt comments, without modifying their sources. If a user chooses to modify source code, the trigger would be automatically removed from source code once the actions are taken. As the result of our design decisions, TrigIt should be integrated in the build process after the compilation step. We also developed a Maven plugin to evaluate the triggers after the compilation phase and prior to testing/packaging phase.

```
1  // AbstractStreamingHasher.java
2  protected AbstractStreamingHasher(int chunkSize, int bufferSize) {
3     // TODO(kevinb): check more preconditions (as bufferSize >= chunkSize)
4     // if this is ever public
5     if (trigItIsPublic())
6        checkArgument(bufferSize >= chunkSize);
7     checkArgument(bufferSize % chunkSize == 0); ... }
8
9  @TrigItMethod
10 boolean trigItIsPublic() {
11    return TrigIt.getMethod("<init>", int.class, int.class).isPublic(); }
```

**Figure 1: An example trigger-action comment from the Google Guava project and TrigIt encoding to illustrate an implicit action.**

We evaluated TrigIt by manually migrating 20 existing trigger-action comments to the TrigIt DSL; all the comments are from 8 popular open-source projects available on GitHub. In our experiments, we report the complexity of TrigIt statements measured in terms of the number of tokens in the specifications and overhead introduced by the tool in the build process. TrigIt does not introduce any overhead at runtime. Our results show that TrigIt introduces negligible overhead in the build process, and triggers and actions are simple to encode.

The main contributions of this paper include:

★ **Idea**: We introduce an idea to encode trigger-action comments, currently written in natural language, as executable statements in the host language. Having executable trigger-action comments enables their maintenance (e.g., refactoring), testing, and automatic evaluation of the triggers and actions when the state of the code repository changes.

★ **Tool**: We implemented our idea in a tool, dubbed TrigIt, for Java. TrigIt statements are written in a language embedded in Java and perform queries and actions over ASTs. TrigIt interposes between compilation and execution, and thus does not introduce any runtime overhead.

★ **Evaluation**: We evaluated TrigIt by manually migrating existing comments, and measuring overhead in terms of the lines of code and execution time during the build process.

## 2 ILLUSTRATIVE EXAMPLES

This section shows several existing trigger-action comments from large open-source projects, the encoding of these comments in the TrigIt DSL, and TrigIt's workflow. We chose the comments such that we can illustrate various aspects of TrigIt.

Figure 1 shows a code snippet from the Google Guava project [14]. The original trigger-action comment is crossed out and our encoding of the executable trigger-action comment is highlighted. This comment was added on commit c92e1c7 (2017-05-31) and is still present in the latest revisions of the project. In this case, a developer wants to add more precondition checks if the method or constructor becomes public. TrigIt requires that the trigger is encoded as a separate method that returns a boolean value, and each TrigIt method needs to have the @TrigItMethod annotation. trigItIsPublic finds the constructor and checks its modifiers. The invocation of trigItIsPublic is a guard for an extra

```
1  // FreemarkerResultMockedTest.java
2  public void testDynamicAttributesSupport() throws Exception { ...
3     dispatcher.serviceAction(request, response, mapping);
4     // TODO : remove expectedJDK15 and if() after switching to Java 1.6
5     String expectedJDK15 = "<input type=\"text\" ...;
6     if (trigItJava6())
7        expectedJDK15 = "<input type=\"text\" ...;
8     String expectedJDK16 = "<input type=\"text\" ...;
9     ...
10    if (trigItJava6())
11       assertEquals(expectedJDK16, result);
12    else
13    if (result.contains("foo=\"bar\" ...))
14       assertEquals(expectedJDK15, result);
15    else
16       assertEquals(expectedJDK16, result); ... }
17
18 @TrigItMethod
19 boolean trigItJava6() {
20    return TrigIt.getJavaVersion().greaterEqualThan(TrigIt.JAVA6); }
```

**Figure 2: An example trigger-action comment from the Apache Struts project and TrigIt encoding to illustrate a trigger based on build configuration.**

```
1  // Mapper.java
2  // TODO: make this protected once Mapper and FieldMapper
3  // are merged together
4  public final String simpleName() {
5     return simpleName;
6  }
7
8  @TrigItMethod
9  public static void checkMerge() {
10    if(!TrigIt.hasClass("Mapper") || !TrigIt.hasClass("FieldMapper")) {
11       TrigIt.getMethod(simpleName()).setProtected();
12    }
13 }
```

**Figure 3: An example trigger-action comment from the Elasticsearch project and TrigIt encoding to illustrate an explicit action.**

precondition check. It is interesting to observe that in this case, the action (transformation step) is implicit, i.e., we simply execute extra statements. As this comment is not specific enough, i.e., we do not know all the preconditions that developers would like to check, we could include an extra action statement to print a warning to developers when the trigger is satisfied. Recall that triggers are evaluated during the build process prior to execution.

Figure 2 shows a code snippet from the Apache Struts project [5], which is a web framework for creating Java web applications. The specified action was performed at commit a5812bf (2015-10-06), five months after the trigger became true. Unlike the previous example, this one illustrates a query statement over the build configuration script. Specifically, the trigger holds if the current Java version is greater than 1.6. Regarding the action, similarly to the previous example, we simply guard several statements that should be removed in this case. As seen in this example, the same TrigIt method can be invoked at several places.

Finally, Figure 3 shows a code snippet from the Elasticsearch project [10], which is a distributed search engine. We use this example to illustrate the explicit code transformation steps. In this example, developers want to change the access modifier of a method (simpleName) from `public` to `protected` if two classes (Mapper and FieldMapper) are merged. Although there is no unique way to encode a trigger that checks if two classes are merged, the check can be approximated in a number of ways. Our approach is to check that one of the classes is no longer available. A better option might be to check that one class is removed while the other one is still present. By knowing the relation between the classes (FieldMapper extends Mapper) and their usage, we believe that the original developers could provide more precise trigger. The action specifies that the modifier of the method should be changed to protected. Unlike prior examples, the action in this case is explicit, and it is expressed as a transformation step over the class AST. It is interesting to observe that triggers and actions find code elements to query or modify in the same style.

## 3 TRIGIT TECHNIQUE

This section describes the TrigIt DSL, presents the workflow of the system and the integration with existing build processes, discusses several phases, and briefly describes the implementation.

### 3.1 Language

**Action types**. We differentiate two types of actions: (1) explicit and (2) implicit. We define an *explicit action* as an action that uses the TrigIt DSL to specify modifications to an AST. These actions modify *out-of-method code elements*, including method signatures, field declarations, class declarations, and others. Figure 3 from the previous section illustrates an explicit action that updates a modifier of a method. We define an *implicit action* as a sequence of statements that should or should not be executed depending on a trigger that guards those statements. We say that these actions modify *in-method* code elements (i.e., statements to be executed). Figures 1 and 2 from the previous section illustrate implicit actions.

**Trigger-action syntax**. The TrigIt specifications are written in a subset of Java with slightly modified semantics as discussed later in this section. Developers have to write all triggers and explicit action statements inside project-unrelated methods annotated with @TrigItMethod; these methods should have neither arguments nor throws clause, but there are no other restrictions on method signatures. The return type of each TrigIt method is either void or boolean; the return type differentiates methods that are used with explicit and implicit actions. Specifically, the following are valid method signatures:

@TrigItMethod void ⟨modifiers⟩ ⟨name⟩();

@TrigItMethod boolean ⟨modifiers⟩ ⟨name⟩();

We require that all TrigIt code be written in separate methods to ensure a clear boundary between executable comments and project code, as well as to enable evaluation of executable comments independently of project code.

TrigIt methods that return a boolean value have to have only a single statement that implements the trigger; the statement has to return the result of a *query expression*. (We discuss the details of the

```
1  package org.trigit.model;
2  public class ClassModel extends ModelBase {
3      public String getName() {...}
4      public Modifiers getModifiers() {...}
5      public Collection<FieldModel> getFields() {...}
6      public Collection<MethodModel> getMethods() {...}
7      ...
8  }
```

**Figure 4: Part of TrigIt's ClassModel API.**

query expressions later in this section.) Namely, the following is a valid template for TrigIt methods that returns a boolean value:

@TrigItMethod boolean ⟨modifiers⟩ ⟨name⟩()
    { return ⟨query expression⟩ }

These TrigIt methods are used by a developer to guard implicit actions. The same TrigIt method can be used to guard any number of statements in same or different methods or classes. For example, a developer can have an if statement to guard a piece of code with a trigger invocation:

if( ⟨trigItMethodName⟩() )
    ⟨statement⟩ else ⟨statement⟩

On the other hand, those methods that do not return any value (i.e., void) are used to encode explicit actions. These methods can have multiple statements, but the first statement has to be an if statement, such that the conditional expression is a query expression. Other statements, which are always a part of then block, are transformation steps on a custom AST. Thus each transformation statement has to start with TrigIt related expression that obtains necessary parts of an AST. Besides TrigIt related code, a developer can use print statements (potentially useful only for debugging). The following is the general format of TrigIt methods with no return value:

@TrigItMethod void ⟨modifiers⟩ ⟨name⟩(){
    if( ⟨query expression⟩ ){
        ⟨action statements⟩ }}

**Query expression**. Each query expression is a logical expression, such that each operand queries the state of the code repository via TrigIt API. The TrigIt API always starts with a method invocation to obtain a *stream* of java files, classes, or build configuration scripts. The elements in a stream are objects that are instances of TrigIt's model. Once a developer obtains a stream, the developer can use any standard Java stream operations [3] (e.g., map, filter, count, etc.) to create a query. In case the stream support is not available in the Java version used by the project, the developers may opt for equivalent operations available in TrigIt. The result of each query has to be a boolean value. Clearly, as arguments to stream operations, the developer can use model classes of java files, classes, build configuration files, fields, methods, etc. to access AST data, including class name, modifiers, return type, and others. TrigIt model classes offer a rich API, which we do not show due to the space constraints. We only show signatures of a few methods from ClassModel in Figure 4, which is a model of a class in the project; many syntactic sugars are available, e.g., get a model of a method by name. For example, we can check if a class is available in the project by executing the following query:

TrigIt.getClasses().findAny("name").isPresent()

$$\text{static } \{...\} \to \bot$$
$$\text{<mod> class ...} \to \bot$$
$$\text{<mod> Type f ...} \to \bot$$
$$\text{<mod> Type m(...) } \{...\} \to \bot, \text{ if @TrigItMethod} \notin \text{<mod>}$$
$$\text{this.f } | \text{ ClassName.f} \to \text{"f"}$$
$$\text{this.m(...) } | \text{ ClassName.m(...)} \to \text{"m"}$$

**Figure 5: Rewrite rules to "prepare" @TrigItMethods for the execution; there is no limit on the number of @TrigItMethods per class.**

Additionally, when constructing queries, a developer can use constant values, as well as field accesses and method invocations; however, the semantics for the latter two differ from the one specified by the Java Language Specification, as we discuss below.

**Action statements**. An explicit action statement, similarly to a query expression, has to start with an invocation of the TrigIt API to obtain a stream of classes. We currently provide only an API for modifying ASTs of classes but not build configuration scripts. Unlike the instances of model classes that are available in the query expressions, the instances of model classes available in actions can be both queried and *modified*, i.e., extra API methods are available to specify modifications. The expressions used as arguments to stream operations may include constants, field accesses, and method invocations. There are no limits on the number of action statements that can be written per TrigIt method. For example, if we want to modify an access modifier of a field "f" in the current class, we can write (in a short form) the following action statement:

`TrigIt.getField(f).setPrivate();`

When writing an implicit action statement, a developer may opt to use an API call available only in new version of a library, e.g., `java.nio.file.Paths#get()`. However, writing such code could result in compilation errors. One approach that the developer can take is to write a method invocation via reflection [4].

**Semantics**. Although evaluation of TrigIt methods closely follows Java semantics, there are two important differences. First, all method invocations and field accesses are substituted with the names of methods and fields, i.e., those methods and fields are never invoked or accessed. In case of a method invocation, all the arguments are ignored. For example, the code snippet above would be modified prior to the evaluation to:

`TrigIt.getField("f").setPrivate();`

We made this decision to avoid using strings to refer to a field, method, or class name unless that is absolutely necessary. Our decision will help to keep comments up-to-date with code, e.g., during refactoring, to avoid what researchers call fragile comments [38].

Second, prior to evaluation of a TrigIt method, the class that contains the method is rewritten to remove anything other than the method. This is done to enable evaluation of the executable trigger-action comments without worrying about the environment that is required to execute any piece of code from the project itself. For example, even loading a class may require substantial setup and execution cost if a static block is present.

To prepare a method for evaluation, we define a set of rewrite rules, shown in Figure 5. Each rule has the following format:

$$before \to after \ [condition]$$

where *before* and *after* are AST elements in Java or an empty string denoted by $\bot$. *condition* defines when a rule is applicable; we omit the condition if the rule always applies. The first four rules remove any AST element from a class that is not a method annotated with @TrigItMethod. The last two rules rewrite each field access and method invocation to the name of the accessed field and invoked method, respectively. We recursively apply the rewrite rules on the method until no more rewrite rules can be applied, the obtained code can be evaluated by standard Java.

**Rationale**. Some of our decisions were discussed in prior paragraphs. We would like to emphasize that our decision to enable the evaluation of trigger-action comments independently on other code was to keep the separation between production code and comments, to enable the evaluation of comments regardless of the requirements needed for running project's code, and to avoid the performance overhead of evaluating comments at runtime. As our decisions are inspired by examples, some of these decisions might need to be revisited in the future to support the encoding of comments as executable statements for a broader class of comments.

### 3.2 Workflow Overview

Figure 6 shows the TrigIt's workflow. TrigIt interposes between the compiler and test execution or the deployment of the project. The first step to use TrigIt is to encode existing trigger-action comments as executable TrigIt statements. Once a project is compiled, the TrigIt methods, and invocations of those methods, are the part of the resulting classfiles. Having executable trigger-action comments checked by the compiler is one advantage over informally written comments.

In the next phase, TrigIt processes all the classfiles from the project to build the AST used by the query expressions and action statements. Next, TrigIt modifies classfiles, based on the rewrite rules in Figure 5, to prepare the TrigIt methods for the execution. The modified classfiles are never stored or disk, unless a developer specifies the debug option, but they are only available in-memory and they are dynamically loaded [31]. TrigIt then evaluates one by one TrigIt method in a non-deterministic order. We discuss potential dependencies between TrigIt methods in Section 5. If the project being built requires a Java version prior to Java 8, which is required for TrigIt execution, TrigIt methods are evaluated via external process invocations.

If a TrigIt method or a query inside a TrigIt method evaluates to true, there are three possible outcomes. First, TrigIt can notify a developer with the list of triggers that hold, without executing any action. In addition to printing which triggers hold, TrigIt also includes a short explanation that justifies the outcome of the trigger (e.g., "Java version modified; file: pom.xml, line: 77"). Second, TrigIt can rewrite the classfiles on disk to ensure that statements guarded by the triggers are executed. In case the trigger is guarding action statements (i.e., void TrigIt method), those statements would be executed in the specified order; no change to a bytecode is performed until all trigger methods are evaluated. Finally, TrigIt can be configured to create a patch for the source code, which developers may inspect and apply. The configuration provided by the user determines what option is taken. The configuration options are not mutually exclusive.

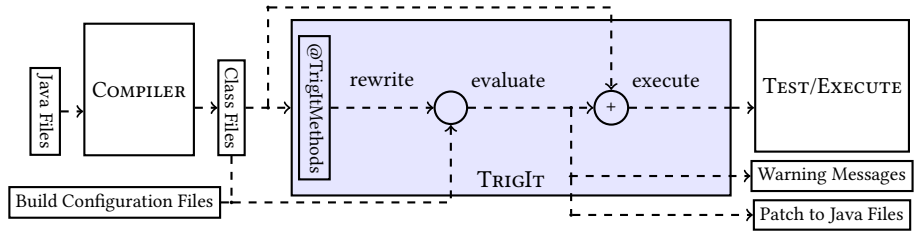

**Figure 6: TrigIt's workflow.**

To ensure that TrigIt methods do not have an impact on the project behavior, all TrigIt methods are removed from the original classfiles after the evaluation of methods is done. For example, if a developer introspects a class to obtain a number of its methods, the presence of TrigIt methods could lead to an incorrect result during program run. Therefore, TrigIt methods are removed.

### 3.3 Implementation

We implemented TrigIt as a standalone Java library that can be used from the command line, or integrated into a build plugin or a testing framework. TrigIt uses the ASM Java bytecode manipulation and analysis framework [8] to rewrite executable code, check correctness of encoding, and transform executable code. More precisely, TrigIt uses the visitor mechanism to build the model classes of the entire project, which are queried with the stream operations. To execute the actions, TrigIt applies the action items, one by one, on the underlying bytecode using ASM.

Additionally, we implemented a Maven plugin to simplify the integration of TrigIt into the build process. We integrated the plugin in the default Maven lifecycle after the compilation phase.

TrigIt provides various options, including "–debug" to show execution steps and store stripped files used to evaluate TrigIt methods, "–assume-true" to force the evaluation of triggers to true, and "–no-action" to tell that no action should be taken. "–assume-true" can be used to check the effect of executing the actions and "–no-action" can be used to check correctness of encoding and report what actions would be taken.

We currently support basic checks of correctness of executable trigger-action comments. Specifically, TrigIt checks if a query expression and action statements refer to code elements that should exist. As an example, consider the following trigger:

`TrigIt.getClass("C").getField("f").isPrivate();`

If class C or field f does not exist, we report an incorrect encoding. These checks are important to detect modifications in code, likely due to software evolution, that invalidate executable comments and notify developers.

Finally, TrigIt can output a patch for the source code if the trigger evaluates to true and actions are executed. Our current approach for creating a patch is a mix of disassembling executable code using jad decompiler and using the knowledge about executed actions. The patch also removes TrigIt method that is satisfied. We expect that a developer would inspect a patch and either apply it or revise code manually.

## 4 EVALUATION

This section describes our evaluation steps. To evaluate TrigIt, we answer the following research questions:

**Table 1: Projects Used in our Experiments, Number of Comments with "TODO" Marker, Number of Potential Trigger-Action Comments (TAC), and Number of Comments We Manually Classified as Trigger-Action Comments ("yes") or Other Type of Comments ("no").**

| Project | SHA | #TODO | #TAC | "yes" | "no" |
|---|---|---|---|---|---|
| amikey/elasticsearch | 850e9d7 | 436 | 94 | 42 | 52 |
| apache/cayenne | 9c07e18 | 380 | 46 | 6 | 40 |
| apache/struts | e2c2ea8 | 62 | 13 | 6 | 7 |
| google/closure-compiler | 3d4f525 | 918 | 171 | 71 | 100 |
| google/guava | ea66419 | 1298 | 161 | 76 | 85 |
| google/j2objc | e85caea | 327 | 71 | 52 | 19 |
| java-native-access/jna | c333527 | 93 | 5 | 2 | 3 |
| jenkinsci/gmaven | 80d5f66 | 28 | 11 | 1 | 10 |
| Avg. | N/A | 442.8 | 71.5 | 32.0 | 39.5 |
| Σ | N/A | 3542 | 572 | 256 | 316 |

**RQ1**: What is the cost of using TrigIt in terms of time overhead introduced in the build process?

**RQ2**: What is the cost of using TrigIt in terms of the number of tokens needed to write the executable trigger-action comments compared to comments written in natural language?

Additionally, to make our dataset complete and enable others to study trigger-action comments, we answer the following question:

**RQ3**: Do trigger-action comments have a common structure that can be learned?

We first describe subjects and todo comments used in the evaluation, and then answer our research questions.

### 4.1 Projects and Comments

Finding trigger-action comments was necessary in our evaluation. In the first step, we selected 8 popular open-source projects. Table 1 shows the list of projects (first column) and SHAs used in our experiments (second column). We selected the projects that differ in size, number of todo comments, and application domain. More importantly, we selected projects based on our prior experience with codebases and build processes. The last requirement was necessary to make the experiments feasible [46]; we wanted projects that we can build to ensure that we can run our tool after migrating the comments. In the second step, we extracted all todo comments from the selected projects. We searched for "TODO" string, which is the most common marker for todo comments [41]. Column 3 in Table 1 shows the number of todo comments for each project.

Table 2: Examples of Encoded Trigger-Action Comments; Examples with Suffix "∗" Only Have Specific Trigger; Examples with Suffix "×N" Indicates the Trigger-Action Comment Occurred N Times in the Project.

| Project | Class | Trigger-Action Comment |
|---|---|---|
| amikey/elasticsearch | Mapper | /** ... make this protected once Mapper and FieldMapper are merged together */ |
| apache/cayenne | DeduplicationVisitor* | // swap inner classes for lambdas when we are on java 8 |
| apache/struts | FreemarkerResultMockedTest | // remove expectedJDK15 and if() after switching to Java 1.6 |
| apache/struts | UIBean×3 | //this is to keep backward compatibility, remove once when tooltipConfig is dropped |
| apache/struts | UIBean* | //use a Boolean model when tooltipConfig is dropped |
| google/closure-compiler | DependencyInfo* | // This would be better as a defender method once Java 8 is allowed (b/28382956): |
| google/guava | AbstractFuture | // investigate using the @Contended annotation on these fields when jdk8 is available |
| google/guava | AbstractStreamingHasher | // check more preconditions (as bufferSize >= chunkSize) if this is ever public |
| google/guava | ClassPath | // Try java.nio.file.Paths#get() when Guava drops JDK 6 support. |
| google/guava | HashCode | // consider ByteString here, when that is available |
| google/guava | SignedBytes | // if Ints.compare etc. are ever removed, *maybe* remove this // one too |
| google/guava | ThrowablesTest | // Remove this guard once lazyStackTrace() works in Java 9. |
| google/j2objc | GeneratedExecutableElement | /** ... Make private when javac conversion is complete. */ |
| google/j2objc | GeneratedTypeElement | /** ... Make private when javac conversion is complete. */ |
| google/j2objc | OptionsTest | // change to 1.9 when Java 9 is supported. |
| java-native-access/jna | IntegerType | // if JDK 7 becomes the min. required use Long#compare(long,long) |
| java-native-access/jna | WinDef | // when JDK 1.7 becomes the min. version, use Boolean.compare(...) |
| jenkinsci/gmaven | SourceDef | // Use URL.toURI() once Java 5 is the base platform |

In the third step, we incorporated knowledge from discourse relations[1] in natural language [40, 51] to look for potential trigger-action comments. We focus on the *condition* relation, that is, the situation in one text segment holds when the condition—as specified in another text segment—is true [27]. In our case, a *condition* relation should hold between each trigger and action. Therefore, we selected todo comments that contain one of the cue words whose presence may signal a *condition* relation within the todo comment: "if", "when", "once", "as", and "then" [35, 36]. Note that in written text, the *condition* relation is almost always signaled by one of the cue words listed above [36]. As the result of the third step, we discovered the total of 572 potential trigger-action comments (fourth column). Note that our selection of comments might have missed some trigger-action comments. However, the selection of comments does not impact the results of our evaluation, as we do not compute any value assuming that the entire set of trigger-action comments is known in advance.

In the fourth step, we manually inspected 572 comments and labeled each comment with "yes" (if the comment is clearly a trigger-action comment) or "no" (if the comment is clearly not a trigger-action comment). Columns 4 and 6 in Table 1 show for each project the number of potential trigger-action comments and the number of these comments with each label. The inspection was done by three authors of this paper together, and in addition to the comment itself, we inspected the context of the comment, i.e., surrounding source code, and potentially other files in the project.

|        | high | medium | low |
|---|---|---|---|
| low    |   4  |   5    | 75  |
| medium |   8  |   9    | 33  |
| high   |  38  |  36    | 48  |

(Action vs Trigger heatmap)

In the fifth step, we inspected 256 trigger-action comments that we annotated with "yes" in the previous step and assigned values to two more labels: specificity_trigger, and specificity_action. Specificity can take one of the following values: (1) "high", which means that we can understand the trigger/action and migration should be feasible, (2) "medium", which means that we mostly understand the trigger/action and migration could potentially be done, and (3) "low", which means that we cannot understand the trigger/action or migration is not feasible. The heatmap above shows the distribution of labels for 256 trigger-action comments. We illustrate the assignment of labels with a couple of examples:

★ *"this is to keep backward compatibility, remove once when tooltipConfig is dropped"* from apache/struts; trigger_specificity: "high", action_specificity: "high".

★ *"When FieldAccess detection is supported, mark that class as reachable there, and remove the containsPublicField flag here"* from google/j2objc; trigger_specificity: "medium", action_specificity: "medium".

★ *"Enable testing for unused fields when ElementUtil glitch is fixed"* from google/j2objc; trigger_specificity: "low", action_specificity: "medium".

★ *"embedded Derby Mode... change to client-server once we figure it out"* from apache/cayenne; trigger_specificity: "low", action_specificity: "low".

Finally, Table 2 shows 20 trigger-action comments that we encoded in TRIGIT and used in our evaluation. Note that one of the comments is used at three different places in apache/struts. We

---

[1]Discourse relations are alternatively called rhetorical or coherence relations.

```
        1            2         3
@TrigItMethod void checkMerge() {
   1       2        3          4           5        6            7
  if( !TrigIt.hasClass("Mapper") || !TrigIt.hasClass("FieldMapper"))
        1            2             3           4
    TrigIt.getMethod("simpleName").setProtected(); }
```

**Figure 7: Example of token counting; we show number of tokens in the structure (first line), trigger (second line), and action (third line).**

used only those comments where trigger_specificity: "high". The table also shows the class that contains the comment, as well as the content of the original comment. Those comments without a specific action are marked with asterisk.

### 4.2 Build Overhead

To compute the time overhead introduced in the build process, we measured the build time (e.g,. mvn test-compile for Maven projects) for each project with and without TrigIt. Recall that TrigIt does *not* introduce any runtime overhead, so we do not run tests during the build, i.e., we report the worst case execution time. If the build is run with TrigIt, we include all the phases, i.e., building the model of the project, evaluating the trigger, and executing the action (see Section 3). We run each configuration five times and report the average time and error; the error is calculated with confidence $p > 0.95$.

We ran experiments on an Intel i7-6700U CPU @ 3.40GHz with 16GB of RAM, running Ubuntu 17.10.

Table 3 shows the results. The first two columns show the project name and the class name that contains the comment(s). Columns 3 and 4 show build time with and without TrigIt, respectively. We can observe that TrigIt introduces *negligible overhead* even for large projects with many classes.

### 4.3 Trigger-Action Comment Complexity

We *estimate* the complexity of writing executable trigger-action comments with the number of tokens needed to encode triggers and actions. Concretely, we count three parts for each executable trigger-action comment: (1) trigger, which is the number of tokens to encode the trigger, (2) action, which is the number of tokens to encode the action, and (3) structure, which is the number of tokens to write necessary boilerplate code, e.g,. method signature. Figure 7 illustrates, using an earlier example, the way we count the tokens. We compare the complexity of executable trigger-action comments with their informal counterparts written in natural language.

Table 3 shows, for each comment, number of tokens in the original comment (column 5), total number (trigger + action + structure) of tokens in the encoded comment (column 6), number of tokens in the trigger (column 7), number of tokens in the action (column 8), and tokens in the structure (column 9)

The results show that even if we count all the tokens in executable trigger-action comments, the increase in the number of tokens compared to the informal text is rather small. If we take into account only the tokens in triggers and actions, executable trigger-action comments are in some cases even shorter than the original comments. Additional benefits of the executable trigger-action comments is that developers can utilize the features of IDEs, such as auto-completion and generation of method signatures, which do not work for comments written in natural language.

### 4.4 Anecdotal Experience

We briefly report on our experience on using TrigIt and interacting with open-source developers to understand their comments. To encode each comment, we read the comment and encoded what we believe were valid triggers and actions. In several cases, we observed that TrigIt always executes the actions, and we thought that there was a bug in the tool. However, by looking at those triggers manually, we found that the triggers (in our opinion) are satisfied but the actions have not been executed. Table 4 shows time when the comment was included in the code repository, time when the trigger is satisfied, and when the action is executed. We only show comments where trigger is satisfied. (Note that Freemarker-ResultMockedTest is a special case, because the developer forgot to remove the todo comment even after the action was taken. We found this todo comment interesting and used older revision in our evaluation.) We reported to the original developers five comments that have satisfied triggers. We got the responses from all just few hours after we submitted the reports. A developer of apache/cayenne immediately performed the action; he also sent a note: "Thank you for the reminder!" A developer of google/j2objc confirmed that the trigger is satisfied and said: "it's time to cleanup the TODOs – any volunteers? :-)". Developers of other comments explained that the comments were not specific enough. For example:

★ "Unfortunately Java 8 still causes some issues with some Google-internal infrastructure. Java 8 is allowed in tests but not in the main part of the code (yet)."

★ "The comment should say something like 'when jdk8 is available to all flavors of Guava.' We currently maintain a backport that targets JDK7 and older versions of Android, and we try to keep the backport and mainline mostly in sync. That's not to say that we couldn't do this, but we'd have to weigh the benefits against the cost of diverging the two."

Some of the responses confirmed the impression that we had while reading todo comments: knowing the details of the project is likely necessary to do valid migration of comments. However, this is not surprising considering that todo comments are written to communicate among developers of the project.

### 4.5 Trigger-Action Comment Structure

Finally, we provide an initial step towards analyzing if trigger-action comments have an underlying structure that may be exploited to automatically identify them. Together with already assigned labels, the initial classifier will provide a good starting point for studying trigger-action comments.

To perform this evaluation step, we used the dataset of todo comments that we manually labeled as described in Section 4.1. In total, we manually annotated 572 todo comments (out of which 256 are confirmed as trigger-action comment).

We designed features and built binary classifiers to automatically identify trigger-action comments from the 572 todo comments. Our classifiers can reach ~80% in accuracy and F-measure, indicating that trigger-action comments can be identified with reasonable reliability; future work can use our results as the baseline.

Table 3: The Effort of Writing and Using the TRIGIT DSL Compared to Using Trigger-Action Comments; The Build Time is Measured 5 Times and Reported as Average(Error), Where the Error is Calculated with Confidence $p > 0.95$.

| | | Build Time (s) | | # Tokens | | | | |
| | | Without | With | Comment | TRIGIT | | | |
| Project | Class | TRIGIT | TRIGIT | | Total | Trigger | Action | Structure |
| --- | --- | --- | --- | --- | --- | --- | --- | --- |
| amikey/elasticsearch | Mapper | 84.52(±1.82) | 87.18(±2.81) | 11 | 14 | 7 | 4 | 3 |
| apache/cayenne | DeduplicationVisitor | 75.75(±2.07) | 77.22(±0.57) | 12 | 21 | 6 | 13 | 2 |
| apache/struts | FreemarkerResultMockedTest | 35.14(±0.31) | 36.60(±0.17) | 12 | 21 | 5 | 2 | 14 |
| apache/struts | UIBean | 18.64(±0.44) | 20.26(±0.14) | 13 | 10 | 3 | 0 | 7 |
| apache/struts | UIBean | 18.61(±0.33) | 20.09(±0.46) | 9 | 16 | 3 | 6 | 7 |
| google/closure-compiler | DependencyInfo | 70.83(±0.57) | 73.38(±0.60) | 22 | 27 | 6 | 19 | 2 |
| google/guava | AbstractFuture | 19.08(±0.73) | 20.98(±0.40) | 15 | 33 | 6 | 24 | 3 |
| google/guava | AbstractStreamingHasher | 18.31(±0.18) | 21.26(±0.44) | 13 | 17 | 8 | 3 | 6 |
| google/guava | ClassPath | 18.58(±0.53) | 21.25(±0.60) | 14 | 22 | 5 | 8 | 9 |
| google/guava | HashCode | 19.81(±0.90) | 22.38(±0.37) | 9 | 15 | 4 | 8 | 3 |
| google/guava | SignedBytes | 18.99(±0.38) | 22.08(±0.31) | 25 | 13 | 6 | 4 | 3 |
| google/guava | ThrowablesTest | 18.80(±0.50) | 21.06(±0.32) | 12 | 12 | 5 | 0 | 7 |
| google/j2objc | GeneratedExecutableElement | 8.88(±0.19) | 11.64(±0.25) | 9 | 17 | 9 | 5 | 3 |
| google/j2objc | GeneratedTypeElement | 9.02(±0.11) | 11.42(±0.30) | 10 | 17 | 9 | 5 | 3 |
| google/j2objc | OptionsTest | 9.25(±0.83) | 11.50(±0.27) | 11 | 18 | 5 | 6 | 7 |
| java-native-access/jna | IntegerType | 1.85(±0.08) | 2.14(±0.11) | 14 | 17 | 5 | 4 | 8 |
| java-native-access/jna | WinDef | 2.62(±0.22) | 3.06(±0.18) | 12 | 17 | 5 | 4 | 8 |
| jenkinsci/gmaven | SourceDef | 7.25(±0.37) | 7.45(±0.27) | 12 | 27 | 5 | 6 | 16 |
| Avg. | N/A | N/A | N/A | 13.1 | 18.6 | 5.7 | 6.7 | 6.2 |

Table 4: The Timeline of the Comments: Added (Column 3-4), Trigger Satisfied (Column 5-6), Action Executed (Column 7-8).

| | | First Added | | Trigger Satisfied | | Action Executed | |
| Project | Class | Commit | Date | Commit | Date | Commit | Date |
| --- | --- | --- | --- | --- | --- | --- | --- |
| apache/cayenne | DeduplicationVisitor | 39b70d1 | 2016-10-02 | b332610 | 2017-08-18 | 87d7e89 | 2018-03-07 |
| apache/struts | FreemarkerResultMockedTest | 0f2c049 | 2012-11-22 | 25cdfd6 | 2015-05-28 | a5812bf | 2015-10-06 |
| google/closure-compiler | DependencyInfo | fc465c1 | 2016-04-25 | 62ba0ab | 2017-10-11 | N/A | N/A |
| google/guava | AbstractFuture | 0b76074 | 2014-11-25 | 86fb700 | 2016-11-04 | N/A | N/A |
| google/guava | ClassPath | 896c51a | 2017-01-12 | 9ebd95a | 2018-02-20 | N/A | N/A |
| google/j2objc | GeneratedExecutableElement | 6eac122 | 2016-12-14 | bc5dbad | 2017-08-31 | N/A | N/A |

**Text processing**. First we normalize each todo comment by removing the prefix "TODO" and any associated meta data (e.g. "TODO (user):"). Then, every todo comment is split to two segments—trigger and action. We use a *condition* discourse relation and hand-craft 10 regular expression templates to extract the two arguments of the relation, which correspond to the "trigger" and the "action". These templates split a todo comment by the *condition* relation cue words, for example: "if *<trigger>*, (then)? *<action>*", "*<action>* once *<trigger>*", "when *<trigger>*, *<action>*".

**Features**. We designed 4 types of features, separately for triggers and actions.

★ *Number of tokens* in the segment, which may capture the amount of information encoded.

★ *Unigram POS-tag* (part of speech tag) *frequency distribution*. We use the coarse-grained POS-tag assigned by spaCy [1]. In total, there are 21 unique POS-tags. This feature characterizes the type of information, e.g., verbs or nouns, in each text segment.

★ *Special token frequency distribution*. We calculate the term frequency distribution for these token classes: stop words, punctuations, number, Java keyword, Java identifier (approximated by word in CamelCase or UPPER_CASE). Stop words and punctuations usually serve syntactic purposes rather than content; the other classes correspond to specific vocabularies used in the source code.

★ *Word embedding*. To capture overall content of a trigger or action, for each token, we convert it to its pre-trained word embedding (a real-valued vector of dimension = 100) from Mikolov

**Table 5: Performance of Automatic Classification Systems for Trigger-Action Comments.**

| System   | Accuracy | F1 Score | Precision | Recall |
|----------|----------|----------|-----------|--------|
| baseline | 0.710    | 0.680    | 0.672     | 0.688  |
| full     | 0.811    | 0.790    | 0.787     | 0.793  |

et al. [26] MetaOptimize dataset [25]. Then we calculate the embedding vector representing the segment as the average of word embedding vectors of all tokens.

**Classification.** We train two classifiers using Logistic Regression algorithm: a baseline, and a full system. The baseline system is trained with all features except word embedding.

The classification systems are evaluated using leave-one-out cross validation. We show the performance of different classification systems in Table 5, measured in terms of accuracy, F1 score, precision and recall. Both systems outperform the random baseline (accuracy = 0.5). The full system reached 0.811 accuracy and 0.790 F1 score. The results provide an initial result that trigger-action comments manifest an underlying structure that can be exploited by machine learning for them to be effectively identified. Furthermore, a compact representation of trigger/action content using word embeddings can be quite helpful to the classifier.

## 5 DISCUSSION

We justify some of our decisions and report on interesting observations during analysis of comments.

**Binary vs. source code analysis.** We carefully considered if our implementation should analyze and modify binaries or source code. As we explained earlier (Section 3), one of the main reasons to analyze binaries was to enable execution/testing of updated code without having an impact on sources.

**Other observations related to comments.** We encountered a large number of interesting cases while reading/analyzing the comments. We describe only three cases here due to the limited space. First, several trigger-action comments that were classified as trigger_specificity: "low" dependent on information available in bug tracking systems. For example: "remove this once #10262 is fixed". We envision an extension of TrigIt that checks the status of open issues by using APIs from bug tracking systems. Second, we found one example that depends on time: "Delete this alias once it has been deprecated for 3 months." Interestingly, this comment is still in the repository even though it has been more than nine months since the comment was added. Supporting a trigger in TrigIt that checks time is trivial. Third, we observed a couple of examples with a trigger that depend on a some simple test case, e.g., "delete this variable and corresponding if statement when jdk fixed java.text.NumberFormat.format's behavior with Float". We are currently extending TrigIt to support these interesting cases.

**Combining comments.** TrigIt does not prevent developers from combining executable and traditional comments. Moreover, a developer can write natural language comment as part of the action that notifies a developer about a satisfied trigger or actions that can be executed. Future work, which can build on our dataset, can develop migration between executable trigger-action comments and traditional comments; the starting point would be work on code summarization and generation from comments [2, 23].

**Future work.** We plan to design a user study to evaluate the complexity of writing executable trigger-action comments. Also, we plan to integrate TrigIt with other build systems and continuous integration services.

## 6 THREATS TO VALIDITY

**External.** We extracted only comments containing "TODO" markers, however, developers use other markers to indicate todo comments, including "FIXME", "XXX", "HACK", etc. Our decision was based on prior work that showed that "TODO" is the most common marker for todo comments [41].

The projects that we used in the evaluation may not be representative of all open-source projects. To mitigate this threat, we used popular open-source projects that are actively maintained and differ in the application domain.

TrigIt support only projects written in the Java programming language. We chose Java because it is one of the most popular languages, and prior work on analyzing (todo) comments showed the need for automating comment maintenance (see Section 7). However, the idea behind the TrigIt is broadly applicable. To check triggers, TrigIt currently requires Java 8 (or later). However, there is no requirement what Java version is used by the project, as long as Java 8 is available on the system.

**Internal.** Our scripts for mining repositories and TrigIt code may contain bugs. We used scripts that were already utilized in prior work and reasoned about the results of those scripts. Additionally, we extensively tested our code.

**Construct.** We initially searched for "TODO" only in the latest available revision at the time we started the experiments. However, our inspection in some cases, as discussed earlier required that we reason about the evolution of comments.

The focus of TrigIt is on triggers and actions related to the content of compiled code and build files. Many comments belong to this category. Support for queries that check data available only in source, e.g., compile time annotations, are left for future work.

## 7 RELATED WORK

We discuss closely related work on (1) comment analysis, (2) code query languages, (3) program transformation, and (4) if-this-then-that recipes.

**Comment analysis.** Ying et al. [53] were among the first to identify the importance and frequency of todo comments. They analyzed ∼200 todo comments in IBM's internal codebase (Architect's Workbench) and categorized those comments in 13 groups. Two of those groups are a subset of trigger-action comments: "communication: self-communication" and "future task: once the library is available...". Fluri et al. [11] studied co-evolution of code and comments. Storey et al. [41] performed a large empirical study of todo comments; their study combined code and comment analysis, interviews with software developers, and a survey of professional software developers. Most importantly, their study found several maintenance issues related to todo comments, and problems with out-of-date todo comments. Haouari et al. [17] performed an empirical study (quantitative and qualitative) on three open-source projects to create a taxonomy of comments; working comments (i.e., todos) were the second most frequent category based on the quantitative study. Ibrahim et al. [19] studied updates of code and comments and found that rare inconsistent updates lead to bugs in future revisions of software. Sridhara [39] developed a rule-based system for identifying out-of-date todo comments; the system uses a combination of information retrieval, natural language processing (NLP), and semantics to check if a comment is up-to-date. Recently, we [28] proposed several techniques for comment and program analysis to support todo comments as software evolves; our motivating study showed a high percentage of trigger-action comments. Pascarella and Bacchelli [32] performed manual classification of comments in several large open-source projects and showed that machine-learning has potential to automate the classification. TrigIt is the first concrete solution to automatically maintain comments (by being able to compile or transform/refactor the executable comments) and code repositories (by automatically executing actions when associated triggers hold). TrigIt is motivated by prior work on todo comments, as our main goal is to simplify maintenance of comments and code, and remove clutter from code.

Our work is related to Ratol and Robillard's work on fragile comments [38]. They introduced a tool for detecting identifiers in comments when refactoring (i.e., renaming) is performed by a developer. TrigIt removes the need for detecting identifiers in comments as many of the trigger-action comments can be encoded as executable Java code, which would be refactored together with other code.

Code validator [42] is a program that maintains dependencies between code and comments. If code changes, code validator notifies a developer that associated comment need to change; all changes need to be performed manually by the developer. TrigIt removes a burden from developers to manually inspect if a comment needs to be updated. Guo et al. [15] proposed comment generation technique to describe design patterns in code. The generated comments are in natural language. Phan et al. [33] proposed a technique to translate documentation from source code and vice versa.

Work on self-admitted technical debt [6, 18, 24, 34, 54] (SATD) identifies an intentional temporary fix. Unlike work on identifying SATD, TrigIt could potentially be used to clean up the codebase when the trigger is satisfied.

Tan et al. [43] designed and developed iComment, which is the first approach for detecting comment-code inconsistencies. iComment automatically extracts implicit program rules from comments and uses those rules to detect inconsistencies. Recent work presented tComment [44] for testing javadoc comments (properties about null values and exceptions). tComment uses NLP to extract properties for a method, and then uses random test generation to test the method and check if the inferred properties hold. Zhou et al. [55] analyzed directives in API comments to detect those comments. Our work is a step towards avoiding inconsistencies between code and trigger-action comments.

**Code query languages**. JQuery [20, 47] and CodeQuest [16] are source code querying tools; the former one uses a logic programming language, while the latter uses Datalog. Recently, Urma and Mycroft [48] proposed source-code queries with graph databases. While prior work mostly targeted program comprehension, TrigIt targets encoding of trigger-action comments with a language embedded in Java. Another difference is that TrigIt supports querying build configuration files.

**Program transformations**. Actions available in TrigIt are closely related to behavior-preserving transformations, i.e., refactorings [12, 29, 30, 45]. Most relevant work is that on scripting refactorings [21, 49, 50], i.e., providing simple building blocks that can be composed in sophisticated transformations. One of the key differences is that TrigIt actions may not be behavior-preserving (e.g., using new API calls or removing a statement). If actions that we discover in the future would require complex code transformations offered by existing refactoring engines, it would be interesting to explore implementing our actions with those engines (which are, unfortunately, frequently tightly integrated with IDEs).

**IFTTT (if-this-then-that) recipe synthesis**. Researchers, including us, have studied synthesizing IFTTT recipes from natural language [7, 9, 22, 37, 52]. IFTTT recipes are short scripts of trigger-action pairs in daily life domains such as smarthome, personal well-being and social networking, shared by users on websites such as IFTTT.com. Frequently a recipe is accompanied by a short natural language description, for example, "turn on your porch lights when the pizza is on its way". In IFTTT recipes, triggers and actions are functions from APIs and services (e.g., Instagram). In our work, triggers and actions are drawn from unstructured TODO comments during software development. Instead of translating from natural language to an already developed target programming language, our goal is to develop the target programming language.

## 8 CONCLUSION

We presented the first approach, dubbed TrigIt, to encode trigger-action comments as executable statements. A developer can use a subset of Java to encode triggers as query statements over abstract syntax trees (ASTs) and build specifications, and actions as transformation steps on ASTs. TrigIt integrates into the build process and interposes between the program compilation and execution. We migrated 20 trigger-action comments from 8 large open-source projects, which inspired the design of TrigIt language. Our evaluation shows that TrigIt introduces negligible overhead in the build process and the number of tokens needed to encode the comments differs only slightly from the original comments written in natural language. Although TrigIt could be improved in several ways, we believe that timely code updates enabled by TrigIt can already have positive impact on code comprehension, maintenance, and developers' productivity.

## ACKNOWLEDGMENTS

We thank Ahmet Celik, Aleksandar Milicevic, Karl Palmskog, Sheena Panthaplackel and Chenguang Zhu for their feedback on this work. This work was partially supported by the US National Science Foundation under Grants Nos. CCF-1652517 and CCF-1704790.

Executable Trigger-Action Comments                                                                                          Conference'17, July 2017, Washington, DC, USA[3] Oracle and/or its affiliates. 2017. Stream (Java Platform SE 8). (2017). https://docs.oracle.com/javase/8/docs/api/java/util/stream/Stream.html.
[4] Oracle and/or its affiliates. 2018. The Reflection API. (2018). https://docs.oracle.com/javase/tutorial/reflect/.
[5] Apache. 2018. Apache Struts. (2018). https://github.com/apache/struts.
[6] Gabriele Bavota and Barbara Russo. 2016. A large-scale empirical study on self-admitted technical debt. In *International Working Conference on Mining Software Repositories*. 315–326.
[7] Shobhit Chaurasia and Raymond J. Mooney. 2017. Dialog for language to code. In *International Joint Conference on Natural Language Processing (Volume 2: Short Papers)*. 175–180.
[8] OW2 Consortium. 2017. ASM home page. (2017). http://asm.ow2.org.
[9] Li Dong and Mirella Lapata. 2016. Language to logical form with neural attention. In *Annual Meeting of the Association for Computational Linguistics*. 33–43.
[10] Elastic. 2018. Elastic Elasticsearch. (2018). https://github.com/elastic/elasticsearch.
[11] Beat Fluri, Michael Wursch, and Harald C. Gall. 2007. Do code and comments co-evolve? On the relation between source code and comment changes. In *Working Conference on Reverse Engineering*. 70–79.
[12] M. Fowler, K. Beck, J. Brant, W. Opdyke, and D. Roberts. 1999. *Refactoring: Improving the Design of Existing Code*. Adison-Wesley.
[13] Apache Gobblin. 2018. Apache Gobblin GitHub Page. (2018). https://github.com/apache/incubator-gobblin.
[14] Google. 2018. Google Guava. (2018). https://github.com/google/guava.
[15] Jhe-Jyun Guo, Nien-Lin Hsueh, Wen-Tin Lee, and Shi-Chuen Hwang. 2014. Improving software maintenance for pattern-based software development: a comment refactoring approach. In *International Conference on Trustworthy Systems and their Applications*. 75–79.
[16] Elnar Hajiyev, Mathieu Verbaere, and Oege de Moor. 2006. CodeQuest: Scalable Source Code Queries with Datalog. In *European Conference on Object-Oriented Programming*. 2–27.
[17] Dorsaf Haouari, Houari Sahraoui, and Philippe Langlais. 2011. How good is your comment? A study of comments in Java programs. In *International Symposium on Empirical Software Engineering and Measurement*. 137–146.
[18] Qiao Huang, Emad Shihab, Xin Xia, David Lo, and Shanping Li. 2018. Identifying self-admitted technical debt in open source projects using text mining. *Empirical Software Engineering* 23, 1 (2018), 418–451.
[19] Walid M. Ibrahim, Nicolas Bettenburg, Bram Adams, and Ahmed E. Hassan. 2012. On the relationship between comment update practices and Software Bugs. *Journal of Systems and Software* 85, 10 (2012), 2293–2304.
[20] Doug Janzen and Kris De Volder. 2003. Navigating and Querying Code Without Getting Lost. In *International Conference on Aspect-oriented Software Development*. 178–187.
[21] Huiqing Li and Simon Thompson. 2012. A domain-specific language for scripting refactorings in Erlang. In *Fundamental Approaches to Software Engineering*. 501–515.
[22] Chang Liu, Xinyun Chen, Eui Chul Shin, Mingcheng Chen, and Dawn Song. 2016. Latent attention for if-then program synthesis. In *Advances in Neural Information Processing Systems*. 4574–4582.
[23] Pablo Loyola, Edison Marrese-Taylor, and Yutaka Matsuo. 2017. A Neural Architecture for Generating Natural Language Descriptions from Source Code Changes. In *Annual Meeting of the Association for Computational Linguistics*. 287–292.
[24] E. D. S. Maldonado, R. Abdalkareem, E. Shihab, and A. Serebrenik. 2017. An empirical study on the removal of self-admitted technical debt. In *International Conference on Software Maintenance and Evolution*. 238–248.
[25] MetaOptimize. 2015. MetaOptimize dataset. (2015). http://metaoptimize.com/.
[26] Tomas Mikolov, Wen-tau Yih, and Geoffrey Zweig. 2013. Linguistic Regularities in Continuous Space Word Representations. In *Proceedings of the 2013 Conference of the North American Chapter of the Association for Computational Linguistics: Human Language Technologies*. 746–751.
[27] Eleni Miltsakaki, Livio Robaldo, Alan Lee, and Aravind Joshi. 2008. Sense annotation in the Penn Discourse Treebank. In *International Conference on Intelligent Text Processing and Computational Linguistics*. 275–286.
[28] Pengyu Nie, Junyi Jessy Li, Sarfraz Khurshid, Raymond Mooney, and Milos Gligoric. 2018. Natural language processing and program analysis for supporting todo comments as software evolves. In *Workshops of the AAAI Conference on Artificial Intelligence*. 775–778.
[29] William F. Opdyke. 1992. *Refactoring object-oriented frameworks*. Ph.D. Dissertation. University of Illinois at Urbana-Champaign.
[30] William F. Opdyke and Ralph E. Johnson. 1990. Refactoring: an aid in designing application frameworks and evolving object-oriented systems. In *Symposium on Object-Oriented Programming Emphasizing Practical Applications*. 145–161.
[31] Oracle. 1999. Chapter 5. Loading, Linking, and Initializing. (1999). https://docs.oracle.com/javase/specs/jvms/se9/html/jvms-5.html.
[32] Luca Pascarella and Alberto Bacchelli. 2017. Classifying code comments in Java open-source software systems. In *International Working Conference on Mining Software Repositories*. 227–237.
[33] Hung Phan, Hoan Anh Nguyen, Tien N Nguyen, and Hridesh Rajan. 2017. Statistical learning for inference between implementations and documentation. In *Software Engineering: New Ideas and Emerging Technologies Results Track*. 27–30.
[34] Aniket Potdar and Emad Shihab. 2014. An exploratory study on self-admitted technical debt. In *International Conference on Software Maintenance and Evolution*. 91–100.
[35] Rashmi Prasad, Nikhil Dinesh, Alan Lee, Eleni Miltsakaki, Livio Robaldo, Aravind Joshi, and Bonnie Webber. 2008. The Penn Discourse TreeBank 2.0. In *International Conference on Language Resources and Evaluation*.
[36] Rashmi Prasad, Eleni Miltsakaki, Nikhil Dinesh, Alan Lee, Aravind Joshi, Livio Robaldo, and Bonnie L Webber. 2007. The Penn Discourse Treebank 2.0 annotation manual. (2007).
[37] Chris Quirk, Raymond J Mooney, and Michel Galley. 2015. Language to Code: Learning semantic parsers for If-This-Then-That recipes.. In *Annual Meeting of the Association for Computational Linguistics*. 878–888.
[38] Inderjot Kaur Ratol and Martin P. Robillard. 2017. Detecting fragile comments. In *Automated Software Engineering*. 112–122.
[39] Giriprasad Sridhara. 2016. Automatically detecting the up-to-date status of ToDo comments in Java programs. In *India Software Engineering Conference*. 16–25.
[40] Manfred Stede. 2011. Discourse Processing. *Synthesis Lectures on Human Language Technologies* 4, 3 (2011), 1–165.
[41] Margaret-Anne Storey, Jody Ryall, R. Ian Bull, Del Myers, and Janice Singer. 2008. TODO or to bug. In *International Conference on Software Engineering*. 251–260.
[42] Adam Svensson. 2015. *Reducing outdated and inconsistent code comments during software development: The comment validator program*. Master's thesis. Uppsala University, Information Systems.
[43] Lin Tan, Ding Yuan, Gopal Krishna, and Yuanyuan Zhou. 2007. /*iComment: bugs or bad comments?*/. In *Symposium on Operating Systems Principles*. 145–158.
[44] Shin Hwei Tan, Darko Marinov, Lin Tan, and Gary T. Leavens. 2012. @tComment: Testing Javadoc comments to detect comment-code inconsistencies. In *International Conference on Software Testing, Verification, and Validation*. 260–269.
[45] Lance Tokuda and Don Batory. 1999. Evolving object-oriented designs with refactorings. In *Automated Software Engineering*. 174–181.
[46] Michele Tufano, Fabio Palomba, Gabriele Bavota, Massimiliano Di Penta, Rocco Oliveto, Andrea De Lucia, and Denys Poshyvanyk. 2017. There and back again: Can you compile that snapshot? *Journal of Software: Evolution and Process* (2017), e1838.
[47] Raoul-Gabriel Urma and Alan Mycroft. 2012. Programming language evolution via source code query languages. In *Workshop on Evaluation and usability of programming languages and tools*. ACM, 35–38.
[48] Raoul-Gabriel Urma and Alan Mycroft. 2015. Source-code queries with graph databases-with application to programming language usage and evolution. *Sci. Comput. Program.* 97, P1 (2015), 127–134.
[49] Mohsen Vakilian, Nicholas Chen, Roshanak Zilouchian Moghaddam, Stas Negara, and Ralph E. Johnson. 2013. A compositional paradigm of automating refactorings. In *European Conference on Object-Oriented Programming*. Berlin, Heidelberg, 527–551.
[50] Mathieu Verbaere, Ran Ettinger, and Oege de Moor. 2006. JunGL: A scripting language for refactoring. In *International Conference on Software Engineering*. 172–181.
[51] Bonnie Webber, Markus Egg, and Valia Kordoni. 2012. Discourse structure and language technology. *Natural Language Engineering* 18, 4 (2012), 437?490.
[52] Pengcheng Yin and Graham Neubig. 2017. A Syntactic Neural Model for General-Purpose Code Generation. In *Annual Meeting of the Association for Computational Linguistics*. 440–450.
[53] Annie T. T. Ying, James L. Wright, and Steven Abrams. 2005. Source code that talks: an exploration of Eclipse task comments and their implication to repository mining. In *International Working Conference on Mining Software Repositories*. 1–5.
[54] F. Zampetti, C. Noiseux, G. Antoniol, F. Khomh, and M. D. Penta. 2017. Recommending when Design Technical Debt Should be Self-Admitted. In *International Conference on Software Maintenance and Evolution*. 216–226.
[55] Yu Zhou, Ruihang Gu, Taolue Chen, Zhiqiu Huang, Sebastiano Panichella, and Harald Gall. 2017. Analyzing APIs Documentation and Code to Detect Directive Defects. In *International Conference on Software Engineering*. 27–37.
11